\documentclass[journal]{IEEEtran}

\ifCLASSINFOpdf
   \usepackage[pdftex]{graphicx}
\else
\fi

\ifCLASSOPTIONcompsoc
 \usepackage[caption=false,font=normalsize,labelfont=sf,textfont=sf]{subfig}
\else
 \usepackage[caption=false,font=footnotesize]{subfig}
\fi

\usepackage{amssymb}
\usepackage{amsmath,amssymb,amsfonts}
\usepackage[nolist]{acronym}
\usepackage{steinmetz} 
\usepackage{cite}
\usepackage{multirow}
\usepackage{enumitem} 
\setlist[description]{
	leftmargin=30pt, 		
	topsep=0pt,             
	itemsep=0pt,            
	labelindent=0pt,
	style=multiline,
}
\usepackage{algorithm2e}

\usepackage[usenames,dvipsnames]{xcolor}

\begin{acronym}
	\acro{IEEE 13NF}{IEEE 13 Node Test Feeder}
	\acro{IEEE 123NF}{IEEE 123 Node Test Feeder}
	\acro{ACT}{Australian Capital Territory}
	\acro{AC}{alternating current}
	\acro{ANU}{Australian National University}
	\acro{CIM}{current injection method}
	\acro{DER}{distributed energy resource}
	\acrodefplural{DER}[DERs]{distributed energy resources}
        \acro{EV}{electric vehicle}
	\acrodefplural{EV}[EVs]{electric vehicles}
	\acro{ESS}{energy storage system}
	\acrodefplural{ESS}[ESSs]{energy storage systems}
	\acro{IEC}{International Electrotechnical Commission}
	\acro{L-QP}{Local-Quadratic Program}
	\acro{LV}{low-voltage}
	\acro{LRVR}{loss reduction per voltage regulated}
	\acro{MV}{medium-voltage}
        \acro{MPC}{model predictive control}
	\acro{OPF}{optimal power flow}
	\acro{OPF-PC}{OPF-based Proportional Control}
	\acro{PCC}{point of common coupling}
	\acro{PF}{power flow}
	\acro{PV}{photovoltaic}
	\acro{QP}{quadratic program}
	\acro{QCQP}{quadratically constrained quadratic program}
        \acro{RHO}{receding-horizon optimization}
	\acro{SOC}{state of charge}
	\acro{TOU}{time-of-use}
	\acro{VB-VVW}{Voltage-Based Volt-Var-Watt}
\end{acronym}

\begin{document}
%

\title{Optimization-Based Control of Distributed \\ Battery Storage in Distribution Networks}

%
%

\author{Wilhiam~de~Carvalho,
        Ahmad~Attarha,~\IEEEmembership{Member,~IEEE,}
        Hemanshu~R.~Pota
\thanks{W. de Carvalho and A. Attarha are with the College of Engineering, Computing and Cybernetics, The Australian National University, Canberra, ACT, 2601 Australia (e-mail: u6916605@alumni.anu.edu.au).}
\thanks{H.~R.~Pota is with School of Engineering and Information Technology, The University of New South Wales, Canberra, ACT, 2601 Australia.}
\thanks{}}

%
%

\markboth{}%
{}
%



\maketitle

\begin{abstract}

We propose a combined global-local control approach to regulate voltage and minimize power losses in distribution networks with high integration of distributed energy resources (DERs). Local controllers embed the fast acting proportional volt-var-watt control law and have their gain (slope) coefficients updated regularly by a global optimization problem at a slower time-scale. Design of optimal coefficients preserve overall system stability and encapsulate inverter and energy limits of controllable DERs. The proposed approach is formulated based on a linear network model (LinDistFlow) and suitable approximations to produce a convex multi-period optimization formulation. Numerical simulations with real-world customer data and two different distribution feeders revealed that our approach provides substantial voltage regulation, while reducing losses by 11 per cent and peak substation power by 26 per cent compared to other state-of-the-art algorithms.

\end{abstract}


%
\IEEEpeerreviewmaketitle

\section{Introduction}

\IEEEPARstart{D}{istribution} networks are undergoing a massive uptake of \acp{DER} such as solar \ac{PV} and \acp{EV}. These \acp{DER} introduce unprecedented variability and technical challenges in distribution grids, including undesirable voltage deviations, power losses and high peak demands \cite{iea2022}. Excess \ac{PV} generation can cause adverse voltage rise, whereas \ac{EV} charging can intensify voltage drop during peak demand and push the grid beyond its operating limits \cite{enea2020}.

Traditional solutions such as on-load tap changers or grid reinforcement present limited efficacy and can be prohibitively expensive \cite{jahangiri2013}. Alternatively, battery storage is an increasingly attractive solution to mitigate those operational challenges, recently showing fast decreasing costs and progressing technology \cite{nrel2019costs}. Small-scale distributed battery storage absorb the excess of \ac{PV} generation during daylight and discharge during periods of peak demand. Proper control algorithms of battery storage substantially enhances the overall operation of distribution networks, reducing adverse voltage deviations, power losses and peak power demands \cite{wang2016}.

Algorithms for controlling numerous battery storage scattered on the distribution grid typically fall into two main categories: system-wide communication-based control \cite{arnold2016, nazir2021, torbaghan2020, robbins2016, joo2017} and local control \cite{decarvalho2020, procopiou2019, decarvalho2023, decarvalho2022appeec, singhal2019, farivar2013, zhou2021, zhu2016}. Communication-based control, e.g. centralized, provide system-wide optimal performance, but solving global optimization problems are insufficiently fast to respond to quick load and generation variations \cite{plytaria2017}. Local (fully decentralized) control responds fast and autonomously to voltage variations, but lacks system-wide visibility and thus often achieve unsatisfactory performance~\cite{alonso2022}. More recently, in an effort to leverage both the high performance of centralized control and the fast response of local controllers, papers have considered a combined central-local (two-layer) control approach \cite{nrel2019der, ranaweera2017, abadi2021, maharjan2021, samadi2014, bidgoli2018, cavraro2018, alonso2022, baker2018, chistyakov2012}.

In \cite{ranaweera2017} local controllers schedule power setpoints of residential batteries to prevent over-voltage caused by excess rooftop solar \ac{PV}. When local controllers are not enough to bring the voltage within thresholds, a central optimization-based controller overwrites the locally-scheduled power setpoints to keep voltages within limits. In \cite{abadi2021,maharjan2021} global optimization problems are regularly solved over the day to provide parameters (rather than power setpoints) to local proportional controllers acting in real time. The local controllers then quickly compute their own reactive power setpoints by simply measuring the local grid voltage. In \cite{samadi2014,bidgoli2018} parameters of local volt-var proportional control are designed by an optimization layer that includes a future time horizon. Proportional control, also referred to as droop control, is a simple and effective strategy for local controllers to tackle voltage deviations, hence recommended by several utilities and grid codes \cite{ausnet2021}. However, proportional gain coefficients must be rigorously designed to prevent voltage oscillations and unstable interactions between inverters. The aforementioned papers lack in analytical design and assessment of overall system stability (convergence) for their proportional gain coefficients.

In \cite{baker2018,chistyakov2012} proportional coefficients of local controllers are regularly designed by a centralized optimization that considers stability constraints. Such studies focus on the actuation of reactive power, using a single-period optimization formulation that does not account for the inter-temporal relationship which is crucial for limited energy storage devices. Other authors \cite{cavraro2018, alonso2022} focused on non-optimal central computation of local gain coefficients. In \cite{cavraro2018} the global computation provides fit-and-forget values, whereas in \cite{alonso2022} a non-optimal central coordinator continually updates the local controllers without the knowledge of a grid model. While non-optimal and model-free approaches might offer simplicity, optimization-based control provides considerably higher performance, facilitates the inclusion of network models and, when cast as a convex problem, presents efficient and reliable solving time.

In summary, we seek to address the aforementioned points with a rigorous design and assessment of stable proportional gains within a multi-period optimization formulation for controlling both real and reactive power of distributed battery storage, while managing energy charge levels.

In more detail, we propose a novel combined central-local control of residential batteries with real and reactive power jointly optimized for effective grid voltage regulation. Local inverters follow the proportional control law, responding fast and autonomously to local voltage measurements, while a global optimization regularly computes and sends updated gain coefficients. Our mathematical formulation fully captures the physics of the grid (i.e. network impedance characteristic) so that real and reactive power are actuated based on their actual effectiveness, and system stability is kept within a safe margin. The global optimization is formulated with a linear power flow model (LinDistFlow) and valid approximations to result in an efficient convex optimization problem. To properly manage the limited energy of residential batteries, the optimization is formulated as a \ac{RHO} problem, similar to a multi-period \ac{OPF}. The proposed control algorithm, termed hereafter as \ac{OPF-PC}, provides significant grid voltage regulation, while effectively reducing grid power losses and peak power demands.

This paper is organized as follows. Section~II presents the mathematical notation and linear model for distribution networks. Section~III presents the control law equations and stability criterion. Section~IV introduces the original non-convex optimization problem and approximations to formulate the convex, multi-period \ac{OPF-PC} approach. Section~V describes the two-layer \ac{OPF-PC} algorithm as a whole. In Section~VI, two benchmark approaches are described. Numerical simulations are presented in Section~VII and conclusions drawn in Section~VIII.

\section{Preliminaries}

\begin{figure}[!t]
	\centering
	\includegraphics[width=0.80\linewidth]{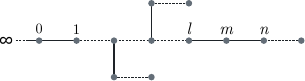}
	\caption{Distribution network represented by a connected graph.}
	\vspace{-0pt}
	\label{fig:graph}
\end{figure}

As in Fig.~\ref{fig:graph}, a radial distribution network is modelled as a connected graph with $1+N$ nodes. Node~0 represents the slack (substation) bus and $N$ is the number of downstream nodes. Let $\mathcal{N}:=\{1,...,l,m,n,...,N\}$ denote the set of all downstream nodes and $\mathcal{E}$ the set of all line segments of the distribution network. For clarity purposes, we focus on the mathematical formulation for a single-phase, with expansion to three-phase notation possible.

\begin{figure}[!t]
	\centering
	\includegraphics[width=0.95\linewidth]{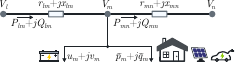}
	\caption{Distribution network notation, with arrows indicating direction of positive power flow. Battery energy storage represents the controllable resource.}
	\label{fig:dn_notation}
\end{figure}

Fig.~\ref{fig:dn_notation} illustrates the notation of the distribution network. Let $r_{lm}$ denote the resistance and $x_{lm}$ the reactance of line segment $(l,m) \in \mathcal{E}$. The real power flowing in the line segment $(l,m) \in \mathcal{E}$ is denoted by $P_{lm}$ and the reactive power by $Q_{lm}$. Voltage magnitude is denoted by $V_m$ and load is divided between controllable and non-controllable load, at each node $m \in \mathcal{N}$. The non-controllable real and reactive load is represented by $\widetilde{p}_{m}$ and $\widetilde{q}_{m}$, respectively. The controllable resource is represented by an inverter-based energy storage device, with real ($u_m$) and reactive ($v_m$) power available to promptly respond to local voltage deviations.

The original power flow equations to model radial distribution networks are complex and non-linear. Optimization problems with such equations are cast as a non-convex problem, which presents poor scalability and solving efficiency and cannot guarantee the global minimizer \cite{borrelli2015}. We consider a linear power flow model as it supports the development of convex optimization problems and facilitates the design of feedback controller gains for overall system stability. Specifically, we consider the established LinDistFlow equations from \cite{baran1989sizing} to model the network within our control approach. This linear model typically provides errors smaller than 1 per cent relative to its non-linear counterpart \cite{farivar2013}. The LinDistFlow equations with our notation in Fig.~\ref{fig:dn_notation} are as follows:
\begin{equation}
P_{lm} = u_{m} + \widetilde{p}_{m} + \sum_{n:(m,n) \in \mathcal{E}} P_{mn},  \ \ \ \ \forall m \in \mathcal{N},
\label{eq:lindist_P}
\end{equation}
\begin{equation}
Q_{lm} = v_{m} + \widetilde{q}_{m} + \sum_{n:(m,n) \in \mathcal{E}} Q_{mn}, \ \ \ \ \forall m \in \mathcal{N},
\label{eq:lindist_Q}
\end{equation}
\begin{equation}
V^2_{m} = V^2_{n} + 2r_{mn} P_{mn} + 2x_{mn} Q_{mn}, \ \ \ \forall (m,n) \in \mathcal{E}.
\label{eq:lindist_E}
\end{equation}
The accuracy and performance of this linear model for designing proportional gains was tested and compared to the exact non-linear model in \cite{decarvalho2022appeec}.
Let $\mathcal{L}_m \subseteq \mathcal{E}$ be the set with line segments of the unique path from node $0$ to node $m$, as in \cite{lin2018,farivar2013}. We then compose a resistance and reactance matrix $\boldsymbol{R},\boldsymbol{X} \in \mathbb{R}^{\text{N}\times \text{N}}$ with their entries obtained as
\begin{align}
R_{i j} = \sum_{(m,n)\in \mathcal{L}_i \cap \mathcal{L}_j} 2r_{mn}, \ \ 
X_{i j} = \sum_{(m,n)\in {\mathcal{L}_i \cap \mathcal{L}_j}} 2x_{mn},
\label{eq:RX_elements}
\end{align}
where $i$ and $j$ represent here the row and column number of the matrices. For each node $m \in \mathcal{N}$, define voltage deviation as $E_m := V_0^2 - V_m^2$, where $V_0$ is the voltage magnitude at node~0 at the reference value of 1~p.u. As in \cite{lin2018,farivar2013}, the LinDistFlow voltage equation is then written for the entire network in matrix format as
\begin{equation}
\boldsymbol{E} = \boldsymbol{R}(\boldsymbol{u} + \boldsymbol{\widetilde{p}}) + \boldsymbol{X}(\boldsymbol{v} + \boldsymbol{\widetilde{q}}),
\label{eq:lin_compact}
\end{equation}
where $\boldsymbol{E}$, $\boldsymbol{u}$, $\boldsymbol{v}$, $\boldsymbol{\widetilde{p}}$, $\boldsymbol{\widetilde{q}} \in \mathbb{R}^{\text{N}}$ are column vectors collecting the variables of each node $m \in \mathcal{N}$ (e.g., $\boldsymbol{u} = [u_1, u_2, ..., u_N]^\top$). Let $\boldsymbol{\widetilde{E}} := \boldsymbol{R} \boldsymbol{\widetilde{p}} + \boldsymbol{X} \boldsymbol{\widetilde{q}}$ be the underlying voltage deviation due to uncontrollable load and rearrange \eqref{eq:lin_compact} as
\begin{align}
\boldsymbol{E} = \boldsymbol{R} \boldsymbol{u} + \boldsymbol{X} \boldsymbol{v} + \boldsymbol{\widetilde{E}}.
\label{eq:lin_compact2}
\end{align}

We define the node-arc incidence matrix $\boldsymbol{D}$ where each row corresponds to a node $m \in \mathcal{N}$ and each column correspond to a line segment $(m,n) \in \mathcal{E}$. Each column $(m,n) \in \mathcal{E}$ in the matrix $\boldsymbol{D}$ has $+1$ in the row $m$, $-1$ in the row $n$ and 0 for the other entries \cite{ahuja1993}. As such, we also write the real and reactive power balance equations \eqref{eq:lindist_P}-\eqref{eq:lindist_Q} in matrix format as:
\begin{align}
\boldsymbol{D} \boldsymbol{P} + \boldsymbol{u} + \boldsymbol{\widetilde{p}} = 0,
\label{eq:lin_compactP}
\end{align}
\begin{align}
\boldsymbol{D} \boldsymbol{Q} + \boldsymbol{v} + \boldsymbol{\widetilde{q}} = 0,
\label{eq:lin_compactQ}
\end{align}
where $\boldsymbol{P}$ and $\boldsymbol{Q}$ are the vectors stacking $P_{mn}$ and $Q_{mn}$, respectively, for every line segment $(m,n) \in \mathcal{E}$.

\section{Proportional Feedback Control}

Droop control actuates power proportionally to local voltage measurements and is a simple and effective strategy to tackle voltage deviations and recommended by different grid codes and standards \cite{samadi2014, alonso2022}. We now formulate the proportional volt-var-watt control for voltage regulation. The classic proportional control law \cite{astrom2020} is used for both real and reactive power, where the linear slope is centered at the reference (nominal voltage). Specifically,
\begin{align}
u_{m}(k) = - \alpha_m E_m(k-1) ,
\label{eq:control_law_u}
\end{align}
\begin{align}
v_{m}(k) = - \beta_m E_m(k-1) ,
\label{eq:control_law_v}
\end{align}
where $k$ is the discrete control step index, $\alpha_m \geq 0$ is the gain coefficient for the controllable real power and $\beta_m \geq 0$ is the gain coefficient for the controllable reactive power. Coefficients are non-negative to ensure local controllers counteract voltage deviations. That is, controllers inject power into the grid to counteract voltage drop and absorb power from the grid to counteract voltage rise. We aim to obtain globally optimal values for $\alpha_m$ and $\beta_m$ for every $m \in \mathcal{N}$.

Recall that $E_m = V_0^2 - V_m^2$ is the voltage deviation and computed by measuring the local voltage $V_m$. Therefore, with given $\alpha_m$ and $\beta_m$ values from an optimization problem, the controllable real and reactive power in \eqref{eq:control_law_u}-\eqref{eq:control_law_v} are locally, autonomously, and quickly computed for voltage regulation. Note that we formulate the general volt-var-watt problem where both controllable real and reactive power are considered. A volt-var control approach considering only controllable reactive power can easily be obtained from this mathematical framework by making $\alpha_m = 0, \ \forall m \in \mathcal{N}$.

In this paper, we focus on the response of inverter-based real and reactive power on the grid operation. The coordination of inverter-based power control with legacy devices has been investigated in \cite{maharjan2021}.

\subsection{System-wide equations: network with feedback control}

Define a diagonal matrix $\boldsymbol{A} \in \mathbb{R}^{\text{N}\times \text{N}}$ in which each diagonal entry is the gain of the respective node as $\boldsymbol{A} := \textrm{diag}(-\alpha_1, ..., -\alpha_N)$. Similarly, define $\boldsymbol{B} := \textrm{diag}(-\beta_1, ..., -\beta_N)$, where  $\boldsymbol{B} \in \mathbb{R}^{\text{N}\times \text{N}}$. Finally, we write the local control law in \eqref{eq:control_law_u}-\eqref{eq:control_law_v} for the entire network as:
\begin{equation}
\boldsymbol{u}(k) = \boldsymbol{A} \boldsymbol{E}(k-1),
\label{eq:contr_law_u_mtrx}
\end{equation}
\begin{equation}
\boldsymbol{v}(k) = \boldsymbol{B} \boldsymbol{E}(k-1).
\label{eq:contr_law_v_mtrx}
\end{equation}
Combining the linear network model \eqref{eq:lin_compact2} with the control law \eqref{eq:contr_law_u_mtrx}-\eqref{eq:contr_law_v_mtrx}, results in
\begin{align}
\boldsymbol{E}(k) = (\boldsymbol{R} \boldsymbol{A} + \boldsymbol{X} \boldsymbol{B}) \boldsymbol{E}(k-1) + \boldsymbol{\widetilde{E}}.
\label{eq:combined_sys_ctrl}
\end{align}
Rearrange \eqref{eq:combined_sys_ctrl} as
\begin{align}
\boldsymbol{E}(k) = \boldsymbol{H} \boldsymbol{G} \boldsymbol{E}(k-1) + \boldsymbol{\widetilde{E}},
\end{align}
where
\begin{align}
\boldsymbol{H} := [\boldsymbol{R} \ \boldsymbol{X}], \ \ \ \ \  \boldsymbol{G} := \begin{bmatrix}
\boldsymbol{A} \\
\boldsymbol{B}
\end{bmatrix}\hspace{-3pt},
\end{align}
with $\boldsymbol{H} \in \mathbb{R}^{\text{N}\times 2\text{N}}$ and $\boldsymbol{G} \in \mathbb{R}^{2\text{N}\times \text{N}}$. Finally, we write the system model with control in the state space format as
\begin{align}
\boldsymbol{E}(k+1) = \boldsymbol{H} \boldsymbol{G} \boldsymbol{E}(k) + \boldsymbol{\widetilde{E}}.
\label{eq:state_space}
\end{align}
As in \cite{baker2018, cavraro2018, singhal2019}, during the fast actuation of electronic power inverters, we consider that the non-controllable load (and thus $\boldsymbol{\widetilde{E}}$) is practically constant. The system in \eqref{eq:state_space} achieves asymptotic stability and convergence to steady-state voltages when
\begin{align}
\rho(\boldsymbol{H} \boldsymbol{G}) < 1  ,
\label{eq:stab_full}
\end{align}
where $\rho(\cdot)$ is the spectral radius defined as the maximum absolute eigenvalue of the matrix. For any value of $\boldsymbol{G}$, in which $\rho(\boldsymbol{H} \boldsymbol{G}) < 1$, the linear system \eqref{eq:state_space} is asymptotically stable and proved to converge to steady-state voltages \cite{singhal2019,cavraro2018,baker2018}.
Since \eqref{eq:stab_full} holds, the system \eqref{eq:state_space} converges to the steady-state equation
\begin{align}
\boldsymbol{E} = \boldsymbol{H} \boldsymbol{G} \boldsymbol{E} + \boldsymbol{\widetilde{E}},
\label{eq:state_space_ss}
\end{align}
and the control equations \eqref{eq:contr_law_u_mtrx}-\eqref{eq:contr_law_v_mtrx} converges to
\begin{equation}
\boldsymbol{u} = \boldsymbol{A} \boldsymbol{E},
\label{eq:contr_law_u_mtrx_ss}
\end{equation}
\begin{equation}
\boldsymbol{v} = \boldsymbol{B} \boldsymbol{E}.
\label{eq:contr_law_v_mtrx_ss}
\end{equation}
In the following section, we design parameters $\boldsymbol{G}$ with the proposed \ac{OPF-PC} approach.

\section{Problem Formulation}

We formulate a global optimization problem to design proportional gain coefficients of all controllable \acp{DER} on the grid. The coefficients are designed to minimize resistive power losses of the entire distribution network, while still providing voltage regulation and stability. The resistive power loss in a line segment $(m,n) \in \mathcal{E}$ is computed as $r_{mn} I_{mn}^2$, where $I_{mn}$ is the current magnitude. We substitute the current magnitude using the complex power equation and formulate the optimization problem as
\begin{equation}
\begin{aligned}
\min_{\substack{\boldsymbol{A}, \boldsymbol{B}}} \quad  &  \sum_{(m,n) \in \mathcal{E}} r_{mn} \frac{(P_{mn}^2 + Q_{mn}^2)}{V_m^2}  \\
\textrm{s.t.}
\quad & \eqref{eq:lin_compact2}, \eqref{eq:lin_compactP}, \eqref{eq:lin_compactQ}, \eqref{eq:stab_full}, \eqref{eq:contr_law_u_mtrx_ss}, \eqref{eq:contr_law_v_mtrx_ss}, \\
\quad & \boldsymbol{A} \leq 0, \\
\quad & \boldsymbol{B} \leq 0. 
\end{aligned}
\label{eq:OPFPC_nonconvex}
\end{equation}
The constraints of optimization problem \eqref{eq:OPFPC_nonconvex} are: network model \eqref{eq:lin_compact2}-\eqref{eq:lin_compactQ}, system stability \eqref{eq:stab_full}, steady-state proportional control law for real \eqref{eq:contr_law_u_mtrx_ss} and reactive \eqref{eq:contr_law_v_mtrx_ss} power and non-negative proportional gains. Coefficients $\boldsymbol{A}, \boldsymbol{B}$ are the only decision variables of interest. Other unknowns of \eqref{eq:OPFPC_nonconvex} are $\boldsymbol{E}, \boldsymbol{P}, \boldsymbol{Q}, \boldsymbol{u}, \boldsymbol{v}$. The optimization problem seeks to minimize resistive power losses in all line segments. Since high power loss is usually caused by periods of peak power demand/export, this objective function also has the effect of flattening the power curve, reducing power peaks and thus preventing congestion in the distribution network. That is, the optimization provides parameters $\alpha_m$ and $\beta_m$ for battery inverters to perform local voltage regulation that also reduces grid power losses and peak demands on the entire network. The objective function of \eqref{eq:OPFPC_nonconvex} and constraints \eqref{eq:stab_full}, \eqref{eq:contr_law_u_mtrx_ss}, \eqref{eq:contr_law_v_mtrx_ss} lead to a non-convex optimization problem. In what follows, we use valid approximations of such equations to formulate a convex optimization problem.

As in \cite{baran1989reconfig,turitsyn2011,sulc2014}, we approximate the loss equation in \eqref{eq:OPFPC_nonconvex} by $r_{mn} (P_{mn}^2 + Q_{mn}^2)$ as voltage magnitude has to be kept around the nominal value ($V_m \approx 1$). With this approximation, the cost function becomes quadratic and convex. The non-convex bilinear equality constraints \eqref{eq:contr_law_u_mtrx_ss}-\eqref{eq:contr_law_v_mtrx_ss} are linearized by a Taylor expansion around the equilibrium point $\boldsymbol{u}=\boldsymbol{v}=0$ and $\boldsymbol{E} = \boldsymbol{\widetilde{E}}$. Such linearization results in $\boldsymbol{u} = \boldsymbol{A} \boldsymbol{\widetilde{E}}$ for the real power and $\boldsymbol{v} = \boldsymbol{B} \boldsymbol{\widetilde{E}}$ for the reactive power. Other options to linearize the bilinear constraints are McCormick relaxations or even Taylor expansion around other equilibrium points, such as the steady-state values of a previous time step.

As in \cite{baker2018}, we represent the stability constraint with the Frobenius norm. Matrix norm is an upper bound on the spectral radius and it holds that $\rho(\boldsymbol{HG})=\rho(\boldsymbol{GH}) \leq \|\boldsymbol{GH}\|_F $. An approximation for the stability constraint is then $\|\boldsymbol{GH}\|_F < 1$. Finally, in order to make it a closed set constraint, the stability constraint is written as $\|\boldsymbol{GH}\|_F \leq 1 - \epsilon$, where $0 < \epsilon \leq 1$ is the stability margin \cite{baker2018}. In general, small $\epsilon$ results in larger gain coefficients, greater voltage regulation and longer convergence to steady-state. Whereas larger stability margins result in smaller gain coefficients, lower voltage regulation and quicker convergence to steady-state \cite{decarvalho2022appeec,farivar2013}. Note that the Frobenius norm boils down to a summation of quadratic terms and thus the stability constraint can be written as a quadratic inequality constraint. Other matrix norms are also possible and are investigated in numerical simulations.

The optimization problem is now formulated as:
\begin{equation}
\begin{aligned}
\min_{\substack{\boldsymbol{A}, \boldsymbol{B}}} \quad  &  \sum_{(m,n) \in \mathcal{E}} r_{mn} (P_{mn}^2 + Q_{mn}^2)  \\
\textrm{s.t.}
\quad & \eqref{eq:lin_compact2}, \eqref{eq:lin_compactP}, \eqref{eq:lin_compactQ}, \\
\quad & \|\boldsymbol{GH}\|_F \leq 1 - \epsilon, \\
\quad & \boldsymbol{u} = \boldsymbol{A} \boldsymbol{\widetilde{E}}, \\
\quad & \boldsymbol{v} = \boldsymbol{B} \boldsymbol{\widetilde{E}}, \\
\quad & \boldsymbol{A} \leq 0, \\
\quad & \boldsymbol{B} \leq 0.
\end{aligned}
\label{eq:OPFPCconvex}
\end{equation}
The optimization problem \eqref{eq:OPFPCconvex} is convex with quadratic objective and quadratic constraints, also known as \ac{QCQP}. Next, we briefly describe the battery storage model and then expand \eqref{eq:OPFPCconvex} to a multi-period optimization to account for energy storage constraints.

\subsection{Multi-Period Optimization}

Let $\mathcal{T}=\{1,2,...,t,...,T\}$ be the steady-state discrete time set, where $t$ is the time index and $T$ is the total number of time steps in the horizon. Each $t \in \mathcal{T}$ has a time length $\Delta(t)$, and time lengths can be different for different $t$.

We model an inverter-based battery storage as the controllable \ac{DER} with limited power and energy. For a residential battery at node $m$, real and reactive power are restricted by the rated apparent power $\overline{s}_m$ as
\begin{equation}
u_m^2(t) + v_m^2(t) \leq \overline{s}_m^2, \ \ \ \ \forall m \in \mathcal{N}.
\label{eq:batt_power}
\end{equation}
Let $\overline{c}_m$ be the maximum, $\underline{c}_m$ the minimum, and $c_m$ the initial energy charge level of a battery storage at node $m$. To model the energy constraints, we first define $\boldsymbol{L}$ by a lower triangular matrix composed of time lengths $\Delta(t)$. Also, define $\boldsymbol{u}_m$ by the vector collecting all steady-state time steps of $u_m(t)$. Specifically,
\begin{align}
\boldsymbol{L} = \begin{bmatrix} 
    \Delta(1)   & 0         &   \dots   & 0         \\
    \Delta(1)   & \Delta(2) &   \dots   & 0         \\
    \vdots      & \vdots    &   \ddots  & \vdots    \\
    \Delta(1)   & \Delta(2) &   \dots   & \Delta(T) 
\end{bmatrix}\hspace{-3pt}, \
\boldsymbol{u}_m = \begin{bmatrix}
u_m(1) \\
u_m(2) \\
\vdots \\
u_m(T) \\
\end{bmatrix}\hspace{-3pt}.
\end{align}
Similar to \cite{ratnam2015net}, we write the upper bound energy constraint as $\boldsymbol{L}\boldsymbol{u}_m \leq \overline{c}_m - c_m$ and lower bound energy constraint as $\boldsymbol{L}\boldsymbol{u}_m \geq \underline{c}_m - c_m$.

We then formulate the \ac{OPF-PC} approach, presented in full here for completeness as
\begin{subequations}
\begin{align}
\min_{\substack{\boldsymbol{A}, \boldsymbol{B}}} \quad  &  \sum_{t \in \mathcal{T}} \sum_{(m,n) \in \mathcal{E}} r_{mn} \big( P_{mn}^2(t) + Q_{mn}^2(t) \big) \Delta(t) \\
\textrm{s.t.}
\quad & \boldsymbol{E}(t) = \boldsymbol{Ru}(t) + \boldsymbol{Xv}(t) + \boldsymbol{\widetilde{E}}(t), \\
\quad & \boldsymbol{D} \boldsymbol{P}(t) + \boldsymbol{u}(t) + \boldsymbol{\widetilde{p}}(t) = 0, \\
\quad & \boldsymbol{D} \boldsymbol{Q}(t) + \boldsymbol{v}(t) + \boldsymbol{\widetilde{q}}(t) = 0, \\
\quad & \|\boldsymbol{G}(t)\boldsymbol{H}\|_F \leq 1 - \epsilon,  \\
\quad & \boldsymbol{u}(t) = \boldsymbol{A}(t) \boldsymbol{\widetilde{E}}(t), \\
\quad & \boldsymbol{v}(t) = \boldsymbol{B}(t) \boldsymbol{\widetilde{E}}(t), \\
\quad & \boldsymbol{A}(t) \leq 0, \\
\quad & \boldsymbol{B}(t) \leq 0, \\
\quad & u_m^2(t) + v_m^2(t) \leq \overline{s}_m^2, \ \ \ \ \forall m \in \mathcal{N},  \\
\quad & +\boldsymbol{L}\boldsymbol{u}_m \leq \overline{c}_m - c_m, \ \ \ \ \ \forall m \in \mathcal{N},  \\
\quad & -\boldsymbol{L}\boldsymbol{u}_m \leq c_m - \underline{c}_m, \ \ \ \ \ \forall m \in \mathcal{N},
\end{align}
\label{eq:OPFPCconvexTH}
\end{subequations}
\hspace{-7pt} where constraints (\ref{eq:OPFPCconvexTH}b)-(\ref{eq:OPFPCconvexTH}j) are for all $t \in \mathcal{T}$. The objective function minimizes the total power losses on the grid over the time horizon (i.e., total energy loss). Constraints (\ref{eq:OPFPCconvexTH}b)-(\ref{eq:OPFPCconvexTH}i) are the same constraints shown before. Constraint (\ref{eq:OPFPCconvexTH}j) represents the apparent power limitation of inverters. Constraints (\ref{eq:OPFPCconvexTH}k)-(\ref{eq:OPFPCconvexTH}l) represent energy limits of battery storage. Problem \eqref{eq:OPFPCconvexTH} is convex and implemented as a \ac{QCQP}.

When a final charge level is desirable at the end of the time horizon, another important constraint to be added in \eqref{eq:OPFPCconvexTH} is $\sum_{t \in \mathcal{T}} \Delta(t) u_m(t) = \hat{c}_m - c_m$, where $\hat{c}_m$ is the final energy charge level. If the final charge level equals to the initial ($\hat{c}_m = c_m$), then the constraint is simplified as $\sum_{t \in \mathcal{T}} \Delta(t) u_m(t) = 0$, which basically describes that the battery has to charge the same amount of energy that discharges over the day.

Note that nodes without battery storage can simply be modelled as a zero power \ac{DER}, i.e., $\overline{s}_m = 0$. Constraint (\ref{eq:OPFPCconvexTH}j) can be modelled by a number of linear inequalities to form a convex polygon constraint, where the original circle is inscribed in the regular n-sided polygon \cite{abadi2021}. In our numerical simulations, we model (\ref{eq:OPFPCconvexTH}j) as a regular octagon to result in a smaller optimization problem and quicker solving time. Voltage bounds are deliberately absent in the optimization problem to avoid infeasibility failures. Due to the stability constraint and limited amount of controllable power and energy, adding voltage bounds to \eqref{eq:OPFPCconvexTH} often results in infeasible problem \cite{abadi2021,zhou2021}. As such, equation (\ref{eq:OPFPCconvexTH}b) can be disregarded to further simplify and speed up the optimization solution.

Unlike standard \ac{OPF} problems, the \ac{OPF-PC} approach in \eqref{eq:OPFPCconvexTH} provides proportional gains $\alpha_m$ and $\beta_m$ for local controllers. Providing gains, rather than power setpoints, enables batteries to compute their own real and reactive power using local voltage feedback. As such, batteries counteract unexpected voltage variations with fast and autonomous response. \ac{OPF-PC} in \eqref{eq:OPFPCconvexTH} also provides gains that consider the specific size (power and energy capacity) of each local controller. This setting prevents irregular participation factors of different batteries across the network, avoiding power saturation and unavailability of controllers due to fully charged/discharged energy storage devices.

\section{Receding-Horizon Optimization}

The optimization in \eqref{eq:OPFPCconvexTH} requires forecast of future uncontrollable load ($\boldsymbol{\widetilde{p}}$ and $\boldsymbol{\widetilde{q}}$). Similar to an \ac{MPC} problem, we solve \eqref{eq:OPFPCconvexTH} in a receding-horizon fashion to account for model and forecast inaccuracies. With the \ac{RHO} method, problem \eqref{eq:OPFPCconvexTH} is solved for a finite time horizon and only the gain coefficients of the first time steps are actually provided to the local controllers. Building on \cite{attarha2020}, we consider a variable time-discretization technique, where time steps further in the horizon are averaged over a longer time length, and time steps closer to the present time have a finer granularity (i.e., $\Delta(1) \leq \Delta(T)$). Since time steps further in the horizon are less precise and less relevant than immediate steps, such a technique significantly reduces the size of the problem and solving time, with negligible impact on performance \cite{attarha2020}.

\begin{figure}[!t]
	\centering
	\includegraphics[width=1.0\linewidth]{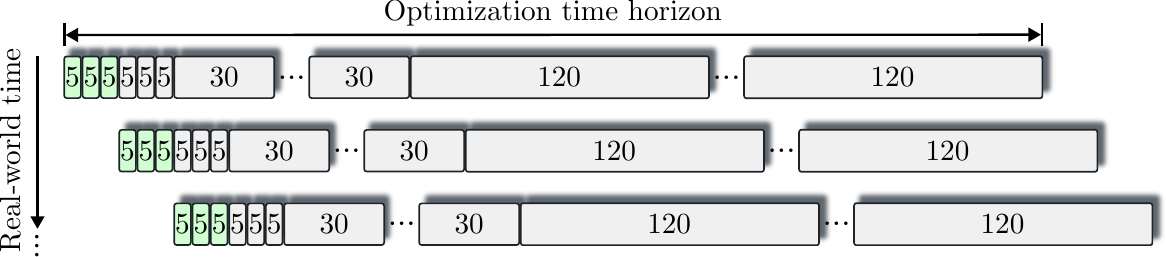}
	\caption{\ac{RHO} with variable steady-state time lengths.}
	\label{fig:MD_RHO}
\end{figure}

Fig.~\ref{fig:MD_RHO} illustrates the \ac{RHO} method with variable time-discretization to solve \eqref{eq:OPFPCconvexTH}. In our numerical simulations, the time horizon is 24~hours: the first six time steps have a 5-min time length ($\Delta(1)=...=\Delta(6)=5/60$~h), the following seven time steps have a 30-min time length ($\Delta(7)=...=\Delta(13)=30/60$~h), and the last time steps have a 2-hour time length ($\Delta(14)=...=\Delta(23)=120/60$~h). Problem \eqref{eq:OPFPCconvexTH} is solved every 15 minutes and gain coefficients of the first three 5-min time steps is actually provided to the local controllers, as illustrated in Fig.~\ref{fig:MD_RHO} in green. Other values for the varying time-discretization technique are also possible.

Algorithm~\ref{alg:one} summarizes the combined central-local (two-layer) \ac{OPF-PC} approach with the \ac{RHO} method. While at the top control layer the central optimization runs at a minute time scale, at the bottom control layer batteries measure and compute their power actions locally as in \eqref{eq:control_law_u}-\eqref{eq:control_law_v} approximately every 100~ms \cite{baker2018,maharjan2021}. Similar to an adaptive control, local controllers have their gain coefficients updated every 5 minutes with globally optimal $\alpha_m$ and $\beta_m$ gains.

\RestyleAlgo{ruled}
\SetKwComment{Comment}{/* }{ */}

\begin{algorithm}[t!]
\caption{\ac{OPF-PC} approach with \ac{RHO}}\label{alg:one}
Get grid matrices $\boldsymbol{R}$, $\boldsymbol{X}$ and $\boldsymbol{D}$; \\
Get battery values $\overline{s}_m$, $\overline{c}_m$, $\underline{c}_m$ for all $m \in \mathcal{N}$; \\
Define stability margin $\epsilon$; \\
\For{every 15 minutes}{
Measure and forecast $\boldsymbol{\widetilde{p}}$ and $\boldsymbol{\widetilde{q}}$; \\
Compute $\boldsymbol{\widetilde{E}}$; \\
Get updated battery charge level $c_m$ for all $m \in \mathcal{N}$; \\
Solve convex optimization problem \eqref{eq:OPFPCconvexTH}; \\
Obtain gain matrices $\boldsymbol{A}(t),\boldsymbol{B}(t)$ for $t \in \{1,2,3\}$; \\
\For{every 5 minutes $(t \in \{1,2,3\})$}{
Update $\alpha_m$ and $\beta_m$ of every local controller; \\
}
}
\end{algorithm}

\section{Benchmark Approaches}

To highlight pros and cons, we present two different control algorithms from the literature to provide grid voltage regulation.

First, we consider the traditional volt-var control, where gain coefficients are obtained by a central coordinator from a non-optimal calculation.

Similar to \cite{cavraro2018,decarvalho2022appeec}, gain coefficients are computed directly by taking $\boldsymbol{A}=\boldsymbol{B}=-g\boldsymbol{I}$, where $\boldsymbol{I}$ is the $\text{N}\times \text{N}$ identity matrix and $g$ is the gain coefficient for every controller. From \eqref{eq:stab_full} and with the addition of the stability margin $\epsilon$, the gain coefficient is computed directly as
\begin{align}
g = \frac{1 - \epsilon}{\rho(\boldsymbol{R}+\boldsymbol{X})},
\label{eq:conv_g}
\end{align}
where $0 < \epsilon \leq 1$ provides asymptotic stability and convergence to steady-state voltages for the system \eqref{eq:state_space}. Note that \eqref{eq:conv_g}, termed Direct approach, provides coefficients that neglect location, size and capacity of controllable \acp{DER}.

The second benchmark approach, termed here as Opt-Bench, is an optimization-based approach to design gain coefficients. Rearranging and approximating the steady-state voltage equation \eqref{eq:state_space_ss} as in \cite{baker2018}, we get
\begin{align}
\boldsymbol{E} = (\boldsymbol{I} - \boldsymbol{HG})^{-1} \boldsymbol{\widetilde{E}} \approx (\boldsymbol{I} + \boldsymbol{HG}) \boldsymbol{\widetilde{E}}.
\label{eq:ss}
\end{align}
Rearrange the approximation as
\begin{align}
\boldsymbol{E} \approx (\boldsymbol{I} + \boldsymbol{R} \boldsymbol{A} + \boldsymbol{X} \boldsymbol{B} ) \boldsymbol{\widetilde{E}}.
\label{eq:ss_approx}
\end{align}

An optimization problem is used to obtain gain coefficients to minimize the maximum voltage deviation in the network ($\|\boldsymbol{E}\|_{\infty}$), while keeping the gains in the stable range \cite{baker2018}. Specifically,
\begin{subequations}
\begin{align}
\min_{\substack{\boldsymbol{A}, \boldsymbol{B}}} \quad  & \|\boldsymbol{E}\|_{\infty}  \\
\textrm{s.t.}
\quad & \boldsymbol{E} = (\boldsymbol{I} + \boldsymbol{R} \boldsymbol{A} + \boldsymbol{X} \boldsymbol{B} ) \boldsymbol{\widetilde{E}}, \\
\quad & \|\boldsymbol{G} \boldsymbol{H}\|_F \leq 1 - \epsilon, \\
\quad & \boldsymbol{A} \leq 0, \\
\quad & \boldsymbol{B} \leq 0.
\end{align}
\label{eq:opt_bench}
\end{subequations}
\hspace{-7pt} The constraints include the steady-state voltage approximate equation (\ref{eq:opt_bench}b), the stability constraint (\ref{eq:opt_bench}c), and non-negative gain coefficients (\ref{eq:opt_bench}d)-(\ref{eq:opt_bench}e). The unknowns of optimization problem \eqref{eq:opt_bench} are $\boldsymbol{A}, \boldsymbol{B}, \boldsymbol{E}$, but gain coefficients are the only decision variables of interest. Problem \eqref{eq:opt_bench} is convex and the objective function can be formulated as a number of linear inequality constraints. Both benchmark approaches provide gain coefficients that do not encapsulate \ac{DER} limitation.

\section{Numerical Simulations}

\begin{figure}[!t]
	\centering
	\includegraphics[width=1.0\linewidth]{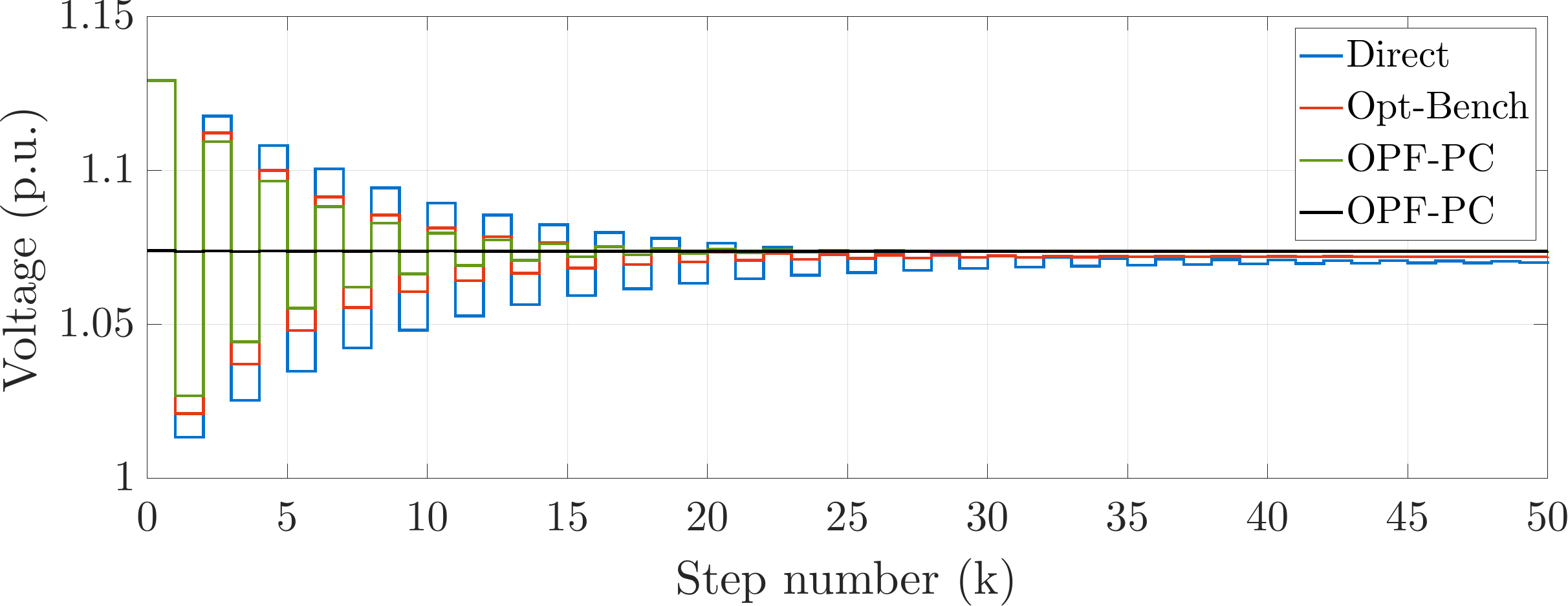}
	\caption{Dynamic voltage behaviour at node~7 for a voltage rise case.}
	\vspace{-0pt}
	\label{fig:dynamic}
\end{figure}

Two different test feeders based on real distribution networks are implemented for numerical simulations. We first consider an 8-node network based on a Belgian residential low-voltage feeder \cite{zeraati2018,mediwaththe2021}. The slack bus is fixed with constant single-phase nominal voltage of 230~V (1~p.u.). We populate the nodes with residential systems with solar \ac{PV} from a real-world, time-varying and de-identified data from the NextGen dataset \cite{shaw2019}. Similar to \cite{decarvalho2022appeec}, two residential systems are connected to every node downstream of the substation bus, with a total of 14 different users on the network. For power flow calculations, loads are modelled as constant power loads and reactive power demand is obtained by considering a power factor of 0.95. Half of the residential systems have a battery storage under the control approach. Stability margin is selected as 10 per cent ($\epsilon = 0.1$), which resulted in a constant proportional gain coefficient of $g=0.45~\text{VA/V}^2$ for the Direct approach in~\eqref{eq:conv_g}. The optimization-based approaches provide tailored gain coefficients for each \ac{DER} and for each 5-min time interval, according to the operating state of the network.

We run the \ac{OPF-PC} approach with the \ac{RHO} method, as described in Algorithm~\ref{alg:one} in Section~V. We carry out simulations on MATLAB R2022a and solve optimization problems with Gurobi Optimizer V10.0.1, using a MacBook Air with Apple M1 2020 chip, 16~GB Ram. To validate the accuracy of the linear model and approximations used in our optimization problem, numerical simulations are run with the full AC non-linear power flow model \cite{garcia2000}. That is, the proposed approach is formulated with the linear model but tested under the full non-linear network model.

\subsection{Dynamic performance}

Similar to \cite{arnold2016} we initially consider unrestricted controllable \acp{DER} to fully assess the dynamic performance of control approaches with abundant controllable power.

Fig.~\ref{fig:dynamic} illustrates the dynamic performance of each approach for an extreme voltage deviation scenario. This extreme scenario represents a case where linear model and approximations are significantly far from exact values. Observe that all approaches achieves overall system stability, converging to steady-state voltages within 50 iterations.

We plot two \ac{OPF-PC} in Fig.~\ref{fig:dynamic} with the same gain coefficients, but with different initial power values. Except for the \ac{OPF-PC} in black line, all local controllers initialize with $u_m(k)=v_m(k)=0$ for $k=0$. Whereas the initial values of the \ac{OPF-PC} in black line are the steady-state values of the previous time step. We observe in Fig.~\ref{fig:dynamic} that proportional controllers provide significant voltage regulation on the grid: from 1.13~p.u. to 1.07~p.u. When the initial controllable power is the steady-state of the previous time step (\ac{OPF-PC} in black line), we observe that voltage converges to steady state almost immediately. Note that all approaches resulted in similar steady-state voltages, as the same stability margin of 10\% was selected. For this numerical simulation with unrestricted controllable \acp{DER}, the stability criterion is the only constraint limiting the increase of proportional gains and consequently reducing voltage deviation.

\begin{figure}[!t]
	\centering
	\includegraphics[width=1.0\linewidth]{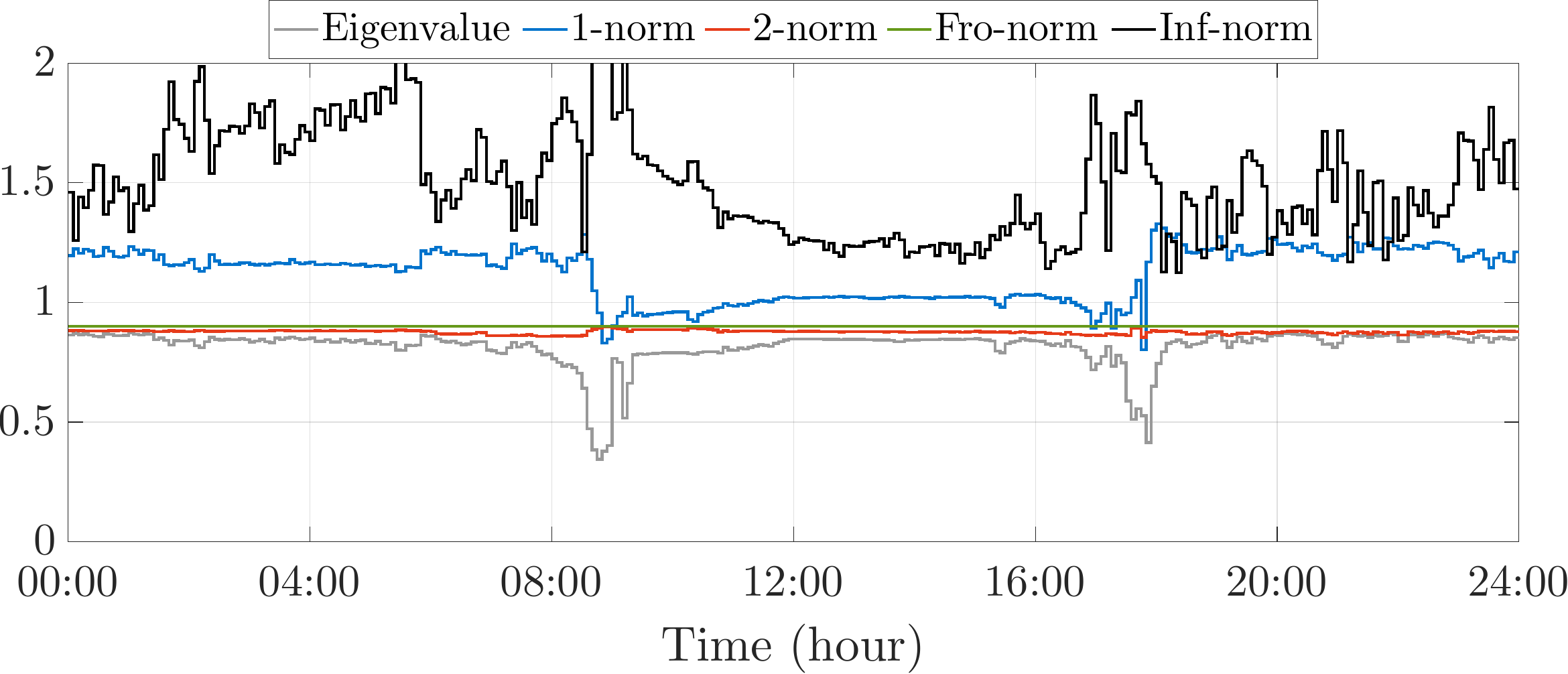}
	\caption{Spectral radius and different norms of $(\boldsymbol{GH})$ for the proposed \ac{OPF-PC} approach over a day.}
	\vspace{-0pt}
	\label{fig:spec_norms}
\end{figure}

Fig.~\ref{fig:spec_norms} illustrates the spectral radius and different norms of $(\boldsymbol{GH})$ as the proportional gains are updated over a day by the proposed \ac{OPF-PC} approach. Observe that the Frobenius norm in green line is at 0.90 throughout the day and is always greater than or equal to the spectral radius in grey line. The spectral radius of $(\boldsymbol{GH})$ represents the dynamics and convergence rate of the system and is clearly less than 1, indicating stability throughout the day. Fig.~\ref{fig:spec_norms} also presents the 1-norm, computed as the maximum absolute column sum of the matrix, Inf-norm, computed as the maximum absolute row sum of the matrix, and 2-norm of matrix $(\boldsymbol{GH})$. The closer the norms are to the spectral radius the closer is the estimate of the actual dynamics of the system and the more precise and less conservative the stability constraint can be. The 2-norm and Frobenius norm are closer to the actual spectral radius. Frobenius norm is simpler to include in optimization problems and results in a convex constraint. The 1-norm and Inf-norm are also simple to compute and include in optimization problems. However, they both are significantly larger than the actual spectral radius and can result in excessively conservative gain values.

\subsection{Steady-state numerical results}

\begin{figure}[!t]
	\centering
	\subfloat[]{%
		\includegraphics[width=1.0\linewidth]{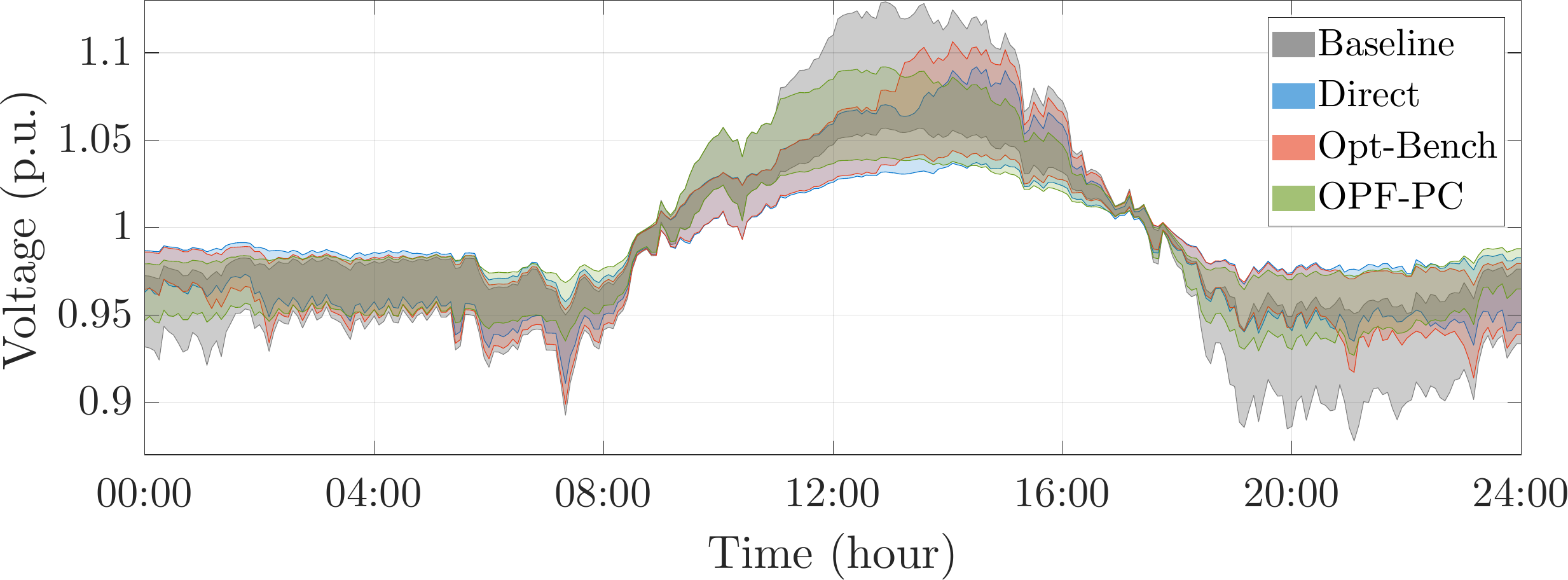}}
	\hfill
	\subfloat[]{%
		\includegraphics[width=1.0\linewidth]{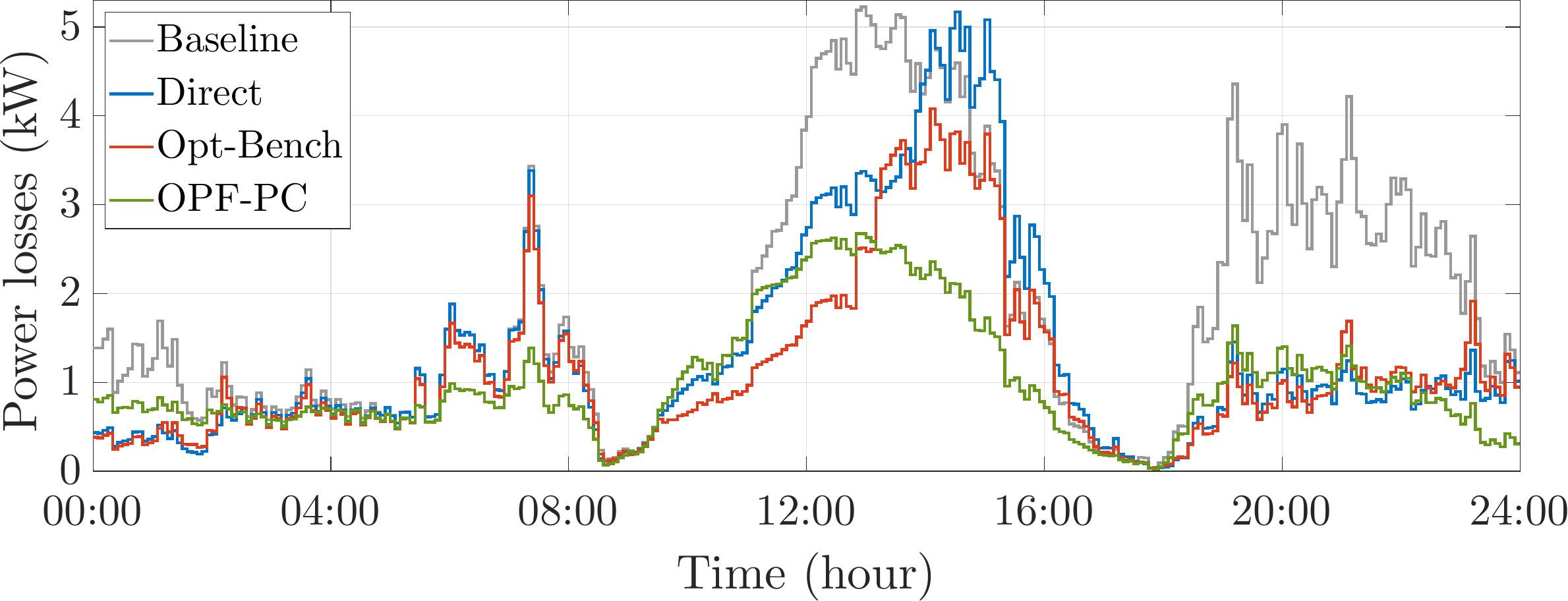}}
	\caption{Steady-state results: (a) Voltage magnitude across the grid and (b) total resistive power losses in the network.}
	\label{fig:ss_venv_loss_c}
\end{figure}

In what follows, we consider that controllable \acp{DER} are similar to a commercial home battery storage. Let the rated apparent power be 5~kVA ($\overline{s}_m = 5$), energy capacity be 10~kWh ($\overline{c}_m=10$ and $\underline{c}_m=0$) and initial charge level be 3~kWh ($c_m=3$). Fig.~\ref{fig:ss_venv_loss_c} illustrates steady-state results for a day with large voltage deviations due to high peak demand and \ac{PV} generation. The range of the voltage magnitude across the network is presented in Fig.~\ref{fig:ss_venv_loss_c}(a) and the total resistive power loss in Fig.~\ref{fig:ss_venv_loss_c}(b). The Baseline simulations represent the no-control case. In Fig.~\ref{fig:ss_venv_loss_c}(a), we observe that grid voltage deviation is significantly reduced when \ac{OPF-PC} is employed, keeping the entire network within the 0.90 and 1.10~p.u. range. In Fig.~\ref{fig:ss_venv_loss_c}(b), the total losses is also considerably reduced when employing the proposed approach, resulting in a substantially more efficient operation of the distribution network.

\begin{figure}[!t]
	\centering
	\subfloat[]{%
		\includegraphics[width=1.0\linewidth]{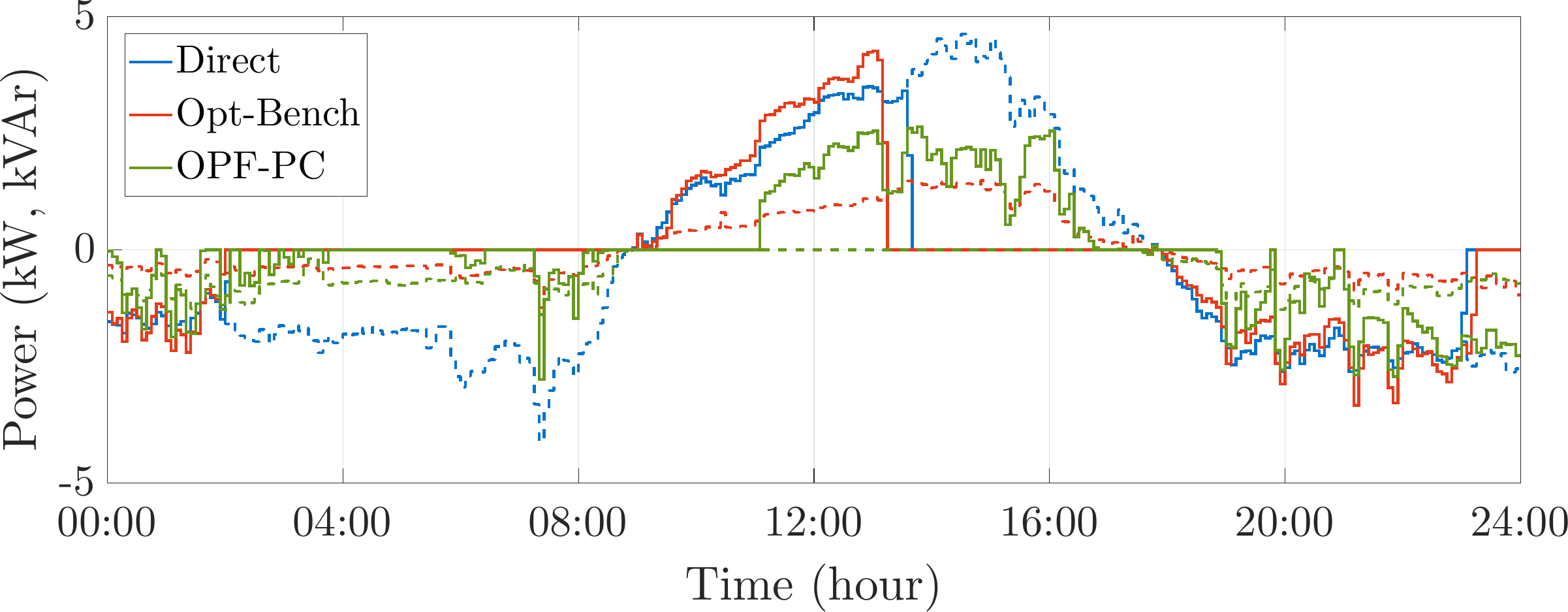}}
	\hfill
	\subfloat[]{%
		\includegraphics[width=1.0\linewidth]{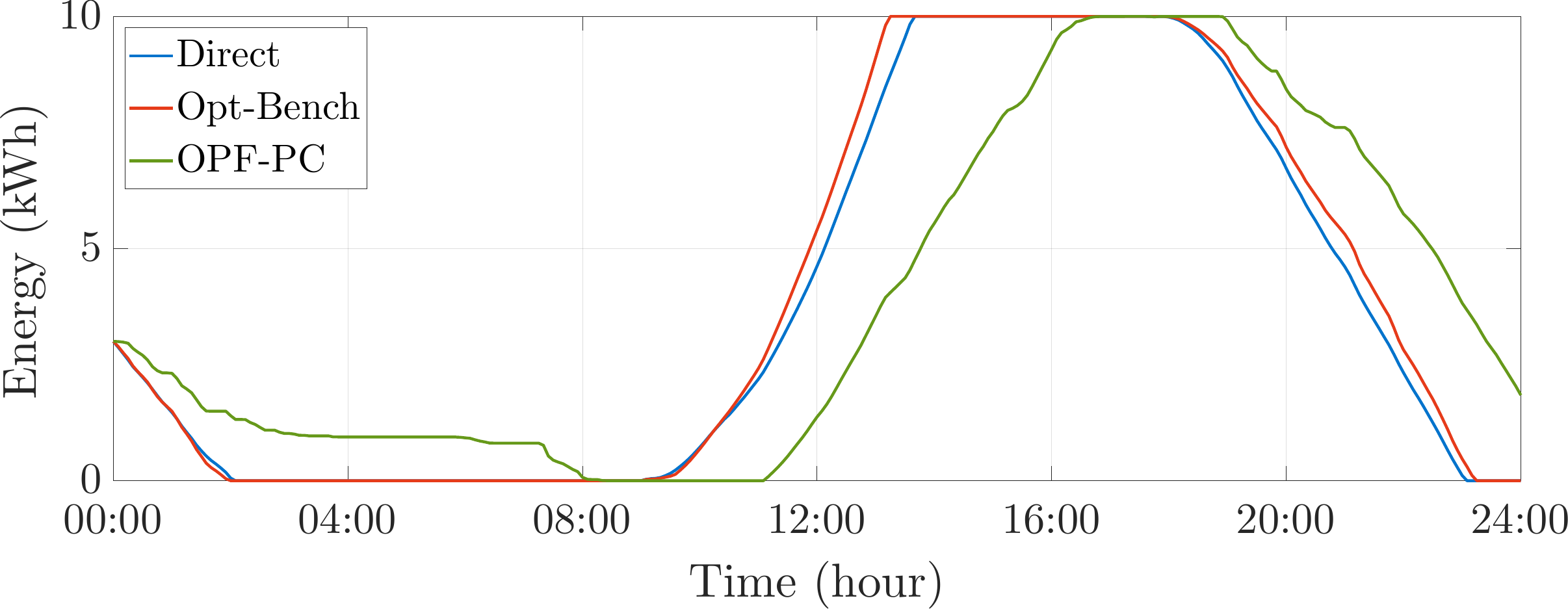}}
	\caption{Controllable \ac{DER} at node~7: (a) real power ($u_m$, solid lines) and reactive power ($v_m$, dashed lines); and (b) energy charge level.}
	\label{fig:ss_uv_ener_c}
\end{figure}

Fig.~\ref{fig:ss_uv_ener_c} illustrates the steady-state controllable real (solid lines) and reactive power (dashed lines) and the energy charge level of the \ac{DER} located at the end of the feeder (node~7). As the Direct and Opt-Bench approaches do not take into account \ac{DER} limitation to design the proportional gain coefficients, the controllable real power is often interrupted due to fully charged/discharged energy levels. With real power unavailable during the critical periods, only reactive power is left to perform voltage regulation. The \ac{OPF-PC} approach encapsulates \ac{DER} power and energy limits into proportional gains and results in real power actuation only during critical times. Observe that \ac{OPF-PC} results in zero $\alpha_m$ and $\beta_m$ gains (thus zero power) during the transition between demand and reverse power flow, when \ac{PV} generation matches power consumption and indeed no voltage regulation is needed. In addition, with the \ac{OPF-PC}, the gains for reactive power control are zero during reverse power flow, resulting in lower grid power losses.

Table~\ref{tab:ss_summary8} presents numerical simulations for other two distinct days of the year with different power consumption and \ac{PV} generation profiles. The results are summarized by four important performance indices: maximum and minimum voltage on the network during that day (Max volt and Min volt), the total energy loss on the network for that day (Ene loss) and the peak apparent power at the substation during that day (S sub). Observe that without controlling \acp{DER} (Baseline), voltage magnitude goes well beyond 1.10 p.u. and 0.90 p.u. All approaches managed to reduce voltage deviation when compared to the Baseline, with the proposed approach providing greater voltage regulation on the network. \ac{OPF-PC} also provided proportional gain coefficients that resulted in local controllers significantly reducing the total energy losses and peak power demand on the network. The considerable reduction in peak power from the substation can prevent overloading transformers and lines and avoid expensive upgrades on the network \cite{nrel2019der}.

\begin{table}[!t]
	\renewcommand{\arraystretch}{1.1}
	\caption{Summary of numerical simulations on the 8-node feeder.}
	\vspace{-5pt}
	\label{tab:ss_summary8}
	\begin{center}
	\begin{tabular}{|c|c|c|c|c|}
		\hline
		\textbf{Index}          & \textbf{Baseline} & \textbf{Direct}   & \textbf{Opt-Bench}    & \textbf{OPF-PC}   \\ \hline
		Max volt (p.u.) 	    & 1.042             & 1.024             & 1.023                 & 1.024             \\ \hline
		Min volt (p.u.)			& 0.886             & 0.904             & 0.893                 & 0.921             \\ \hline
		Ene loss (kWh)		    & 24.3              & 22.2              & 20.5                  & 17.7              \\ \hline
		S sub (kVA)             & 44.9              & 44.3              & 42.7                  & 30.7              \\ \hline \hline
		Max volt (p.u.)         & 1.137             & 1.111             & 1.125                 & 1.099             \\ \hline
		Min volt (p.u.)			& 0.932             & 0.962             & 0.964                 & 0.963             \\ \hline
		Ene loss (kWh)			& 30.6              & 27.8              & 21.4                  & 18.5              \\ \hline
		S sub (kVA)				& 53.3              & 57.7              & 53.9                  & 39.5              \\ \hline
	\end{tabular}
	\end{center}
\end{table}

\begin{table}[!t]
	\renewcommand{\arraystretch}{1.1}
	\caption{Summary of numerical simulations on the 42-node feeder.}
	\vspace{-5pt}
	\label{tab:ss_summary42}
	\begin{center}
	\begin{tabular}{|c|c|c|c|c|}
		\hline
		\textbf{Index}          & \textbf{Baseline} & \textbf{Direct}   & \textbf{Opt-Bench}    & \textbf{OPF-PC}  \\ \hline
		Max volt (p.u.) 	    & 1.004	            & 1.004	            & 1.003	                & 1.004            \\ \hline
		Min volt (p.u.)			& 0.930	            & 0.952	            & 0.957	                & 0.958            \\ \hline
		Ene loss (MWh)		    & 3.52	            & 2.83	            & 2.79	                & 2.63             \\ \hline
		S sub (MVA)             & 12.84	            & 11.77	            & 11.72	                & 9.79             \\ \hline \hline
		Max volt (p.u.)         & 1.039	            & 1.028	            & 1.025	                & 1.034            \\ \hline
		Min volt (p.u.)			& 0.969	            & 0.980	            & 0.982	                & 0.979            \\ \hline
		Ene loss (MWh)			& 5.96	            & 5.32	            & 5.68	                & 4.50             \\ \hline
		S sub (MVA)				& 19.69	            & 18.85	            & 19.44	                & 17.24            \\ \hline
	\end{tabular}
	\end{center}
\end{table}

We now scale up the size of numerical simulations to include over 5500 residential customer on a 42-node distribution feeder based on a medium-voltage network from Southern California Edison \cite{zhou2021}. We populate the network with de-identified customer data from the real-world, time-varying NextGen dataset \cite{shaw2019}. Half of the customers have a controllable battery storage with the same size as described before and stability margin is selected as 10 per cent ($\epsilon = 0.1$). 

Table~\ref{tab:ss_summary42} summarizes numerical simulations on the 42-node feeder for two distinct days of the year with different power consumption and \ac{PV} generation profiles. Observe that all approaches brought the voltage closer to the nominal when comparing to the Baseline, despite milder voltage deviations on the network. The voltage slightly closer to the nominal for the benchmark approaches on the second day is due to unnecessary control of reactive power that results in excessive grid power losses and high power peaks, particularly during solar \ac{PV} exports. \ac{OPF-PC} have shown solid reduction in grid power losses and peak power from substation in all cases. Minimizing grid power losses, as in \ac{OPF-PC}, have clearly contributed to flatten peak power demand/export and to prevent congestion in distribution networks.

Table~\ref{tab:vars_time} presents the number of variables and constraints in the optimization problem of both optimization-based approaches for both test feeders. The solving time to output gain coefficients for the next 15~min is also presented in Table~\ref{tab:vars_time}. Note that the Direct approach is not optimization-based and thus presents negligible central computation requirement. Mainly due to the future time-horizon capability, the proposed approach presents larger problem size and computational time. Even with the conventional personal computer used for numerical simulations in this paper, the \ac{OPF-PC} computation time is much smaller than the updating period of gain coefficients. That is, for the larger network with 5500 residential customers, the proposed approach took 4.42s out of the 15-minute time window available to solve the global optimization problem.

\begin{table}[!t]
	\renewcommand{\arraystretch}{1.1}
	\caption{Problem size and computational time.}
	\vspace{-5pt}
	\label{tab:vars_time}
	\begin{center}
	\begin{tabular}{|c|c|c|c|c|}
		\hline
		\textbf{Approach}                     &\textbf{Size}  & \textbf{\# Variables}     & \textbf{\# Constraints}     & \textbf{Time (s)}    \\ \hline
		\multirow{2}{*}{\textbf{Opt-Bench}}   & 8 nodes       & 22                        & 22                          & 0.03                 \\ \cline{2-5}
	    		                             & 42 nodes      & 124                       & 156                         & 0.07                 \\ \hline
            \multirow{2}{*}{\textbf{OPF-PC}}	  & 8 nodes       & 966                       & 2277                        & 0.34                 \\ \cline{2-5}
            		                           & 42 nodes      & 5658                      & 13225                       & 4.42                 \\ \hline
	\end{tabular}
	\end{center}
\end{table}

\section{Conclusion}

We propose an optimization-based approach to design coefficients of local proportional volt-var-watt controllers to provide voltage regulation in distribution networks. The gain coefficients are designed to reduce total grid power losses, while considering \ac{DER} constraints and keeping overall system stability. \ac{OPF-PC} is solved in a receding-horizon fashion and gains are provided regularly to local controllers, which then act fast and autonomously to voltage variations. Numerical simulations with the full AC non-linear power flow model have shown convergence to steady-state voltages and greater performance when local controllers are tuned by the \ac{OPF-PC} approach. With longer yet viable computational time, the \ac{OPF-PC} approach results in substantial grid voltage regulation and solid reduction of grid power losses and peak demand compared to the state-of-the-art.



\ifCLASSOPTIONcaptionsoff
  \newpage
\fi



\bibliographystyle{IEEEtran}
\bibliography{./bibtex/mybibfile}

\end{document}